\documentclass[pra,twocolumn,aps,amssymb,showpacs,superscriptaddress]{revtex4-1}

\usepackage{epsfig}
\usepackage{graphicx}
\usepackage{amsmath}
\usepackage{amssymb}
\usepackage{enumerate}
\usepackage{hyperref}

\usepackage{color}
\definecolor{rosso}{rgb}{1,0,0}
\definecolor{verde}{rgb}{0,1,0}
\definecolor{blue}{rgb}{0,0,1}
\definecolor{verdescuro}{rgb}{0,0.5,0.5}
\definecolor{rossoscuro}{rgb}{0.7,0.3,0}
\definecolor{bluscuro}{rgb}{0.3,0,0.7}
\definecolor{magenta}{rgb}{1,0,1}

\begin{document}

\title{Bound states in a superfluid vortex: A detailed study along the BCS-BEC crossover}

\author{S. Simonucci}
\affiliation{School of Science and Technology, Physics Division, Universit\`{a} di Camerino, 62032 Camerino (MC), Italy}
\affiliation{INFN, Sezione di Perugia, 06123 Perugia (PG), Italy}
\author{P. Pieri}
\email{pierbiagio.pieri@unicam.it}
\affiliation{School of Science and Technology, Physics Division, Universit\`{a} di Camerino, 62032 Camerino (MC), Italy}
\affiliation{INFN, Sezione di Perugia, 06123 Perugia (PG), Italy}
\author{G. Calvanese Strinati}
\email{giancarlo.strinati@unicam.it}
\affiliation{School of Science and Technology, Physics Division, Universit\`{a} di Camerino, 62032 Camerino (MC), Italy}
\affiliation{INFN, Sezione di Perugia, 06123 Perugia (PG), Italy}
\affiliation{CNR-INO, Istituto Nazionale di Ottica, Sede di Firenze, Firenze 50125 (FI), Italy}


\begin{abstract}
The bound states that can occur in a superfluid vortex have recently called for attention owing to the capability of detecting them experimentally.
However, a detailed theoretical account for the presence of these vortex bound states is still lacking, for all temperatures in the superfluid phase and couplings along the BCS-BEC crossover.
Here, we fill this gap and present a systematic theoretical study based on the Bogoliubov-de~Gennes equations for the bound states that occur over the two characteristic (inner and outer) spatial ranges in which the extension of a superfluid vortex can be partitioned.
It is found that the total number of bound states decreases from the BCS (weak-coupling) side of the crossover toward the intermediate-coupling region where they are still present, whereas the bound states disappear upon entering the BEC (strong-coupling) side.
A scaling relation is also obtained that connects the number of bound states in the inner spatial range of the vortex to the depth and width of the vortex itself.
A criterion is finally provided in terms of the local density of states, to distinguish where a given fermionic superfluid is located in the coupling-temperature phase diagram of the BCS-BEC crossover.
\end{abstract}

\maketitle

\section{Introduction} 
\label{sec:introduction}

Recently, interest has arisen in the internal structure of vortices in type-II superconductors, owing to the presence of bound states in the vortex core region of a Y123 superconductor that were detected experimentally \cite{Berthod-2017}.
The relevance of this finding has been highlighted \cite{SX-2017}, as an indication that the superconducting state of high-temperature superconducting materials should be well described by the conventional BCS pairing theory \cite{BCS-1957}, in spite of the fact that the normal state of these materials is highly unconventional.

In Ref.~\cite{Berthod-2017} the experimental findings were supported by theoretical calculations, based on the approach to the solution of the Bogoliubov-de~Gennes (BdG) equations \cite{DeGennes-1966} developed in Ref.~\cite{Berthod-2016}, to come up with the local density-of-states as the relevant quantity to be compared with the experimental data.
Although these calculations considered the effects of a realistic band structure as well as the influence of disorder and nearby vortices on a
given vortex, they were limited to zero temperature and to a single value of the inter-particle coupling.
To understand in a more complete fashion the properties of bound states in superconducting vortices, it would then be desirable to characterize them also as a function of temperature and inter-particle coupling, at least for the simplest case of an isolated vortex.
Purpose of this paper is to address this aspect of the problem, thereby complementing to some extent the theoretical information provided in Ref.~\cite{Berthod-2017}.

The occurrence of bound states in a superconducting vortex was originally proposed in Ref.~\cite{CdGM-1964} by solving the BdG equations.
That calculation was restricted to zero temperature and to the case of weak coupling for which $\Delta_{0} \ll E_{F}$, where
$\Delta_{0}$ is the bulk value of the gap parameter away from the center of the vortex and $E_{F}$ the Fermi energy.
In more modern language, this would correspond to the BCS (weak-coupling) limit of the BCS-BEC crossover \cite{Physics-Reports-2018}.
Later works, that considered the self-consistent solution of the BdG equations for a vortex line, have either relaxed the restriction to zero temperature but still at weak coupling \cite{Schluter-1991}, or spanned the whole BCS-BEC crossover but still at zero temperature \cite{Sensarma-2006,Levin-2006}.
To obtain a more complete information on the properties of bound states in a superconducting vortex, however, it would be worthwhile to address both their coupling \emph{and\/} temperature dependences at the same time. 

From a technical point of view, in Refs.~\cite{Schluter-1991,Sensarma-2006,Levin-2006} the study of a single vortex was performed in terms of the
BdG equations by constraining the vortex into a cylinder of radius $R_{\mathrm{c}}$ with infinite walls, such that all single-particle wave functions were to vanish at $R_{\mathrm{c}}$.
In this way, however, all single-particle energies turn out to be discrete, in such a way that no sharp distinction there exists between truly bound states with finite spatial extension and unbound continuum states past a well-defined threshold.
This limitation was later removed in Ref.~\cite{SPS-2013}, where the BdG equations were solved by introducing suitable Òfree boundary conditionsÓ through which the vortex profile is matched to its asymptotic behavior at a large distance from the center.
In this way, a clear distinction can be marked between bound (localized) states and unbound states extending to infinity.
By this approach, the temperature dependence of the healing length $\xi$ associated with the size of the vortex could be determined even for temperatures quite close to the critical temperature $T_{c}$ at which 
$\xi$ diverges.
In Ref.~\cite{SPS-2013} this analysis was carried out from weak to strong coupling across the whole BCS-BEC crossover, to determine not only the profile of the gap parameter but also those of the local number 
density and current.
In addition, in Ref.~\cite{SPS-2013} an advanced regularization procedure was developed for the self-consistent gap equation, which permits to increase considerably the accuracy of the numerical calculations.

In the present paper, we take advantage of the above procedures developed in Ref.~\cite{SPS-2013} and concentrate our analysis on the bound-state part of the spectrum for an isolated vortex embedded in an otherwise homogeneous fermionic superfluid of infinite extent.
In this context, we will determine the \emph{local density of states} as a function of energy and of the spatial position from the vortex core, for varying temperature in the superfluid phase from $T=0$ to $T=T_{c}$ and for couplings along the BCS-BEC crossover.
We will thus also be able to study how the total number of bound states evolves as a function of temperature and coupling, paying additional attention to how the bound states are distributed over the two characteristic spatial ranges in which the vortex (including its tail) can be partitioned.

The main results obtained in this paper are as follows:

\noindent
(i) By relying on the numerical procedure to solve the BdG equations for a fermionic superfluid vortex developed in Ref.~\cite{SPS-2013}, 
an energy threshold is identified that clearly separates truly bound states from continuum states.
In this way, the counting of bound states for given coupling and temperature becomes a meaningful process.

\noindent
(ii) In addition, by partitioning the spatial extent of the vortex into an inner and an outer region (to be identified below), the number of bound 
states that hinge on each region is determined as a function of energy below threshold.
It is found that the number of bound states in the inner region scales in a universal way on the depth and width of the vortex itself.

\noindent
(iii) The number of bound states is found to be non-negligible even in the intermediate-coupling region of the crossover, where the Cooper pair size becomes comparable with the inter-particle distance.
To the extent that this small size is compatible with the occurrence of high-temperature superconductivity \cite{Dagotto-1994}, there is thus no a priori reason to associate the occurrence of bound states in a vortex with the superconductivity being of the conventional BCS type as recently asserted \cite{SX-2017}.

\noindent
(iv) Finally, it is proposed that the shape of the local density of states vs energy, taken about the center of the vortex, can serve as a guide to distinguish where a given fermionic superfluid stands
in the coupling-temperature phase diagram of the BCS-BEC crossover.
In this context, one should recall that the local density of states has accurately been measured with a scanning tunneling microscope also in high-temperature superconductors \cite{RMP-2007}, as well as in more complex SNS structures made with conventional superconductors \cite{Esteve-2008}.

The plan of the paper is as follows.
Section \ref{sec:formal_aspects} recalls the main features of the solution of the BdG equations drawn from Ref.~\cite{SPS-2013} for an isolated vortex embedded in an infinite superfluid.
Emphasis is there given to the occurrence of an inner and outer region with different spatial behaviors of the vortex profile where the bound states can reside.
Section \ref{sec:numerical_results} presents the numerical results for the number of bound states and the local density of states, from which
the scaling relation and the criterion for spotting the position along the BCS-BEC phase diagram mentioned above are derived.
Section \ref{sec:conclusions} gives our conclusions.
Appendix \ref{sec:appendix} discusses the asymptotic profile of an isolated vortex in the context of the Gross-Pitaevskii equation, which holds for the composite bosons that form in the BEC limit of the crossover.

\section{Formal aspects} 
\label{sec:formal_aspects}

We begin by briefly recalling the method used in Ref.~\cite{SPS-2013} to solve the BdG equations in the presence of an isolated vortex embedded in an otherwise infinite homogeneous superfluid, as a function of both coupling across the BCS-BEC crossover and temperature.

The BdG equations read \cite{DeGennes-1966}:
\begin{equation}
\left( 
\begin{array}{cc}
\mathcal{H}(\mathbf{r}) & \Delta(\mathbf{r})            \\
\Delta(\mathbf{r})^{*}  & - \mathcal{H}(\mathbf{r})  
\end{array} 
\right)
\left( \begin{array}{c}
u_{\nu}(\mathbf{r}) \\
v_{\nu}(\mathbf{r}) 
\end{array} 
\right) 
= \varepsilon_{\nu}
\left( \begin{array}{c}
u_{\nu}(\mathbf{r}) \\
v_{\nu}(\mathbf{r}) 
\end{array} 
\right)    \, .                                     \label{BdG-equations} 
\end{equation}
\noindent
Here, $\mathcal{H}(\mathbf{r}) = - \nabla^{2}/2m - \mu$ where $m$ is the fermion mass and $\mu$ the chemical potential (we set $\hbar = 1$ troughout).
The local gap parameter $\Delta(\mathbf{r})$ in Eq.~(\ref{BdG-equations}) is determined via the self-consistent condition:
\begin{equation}
\Delta(\mathbf{r}) = - v_{0} \sum_{\nu} u_{\nu}(\mathbf{r}) v_{\nu}(\mathbf{r})^{*} \left[ 1 - 2 f_{F}(\varepsilon_{\nu}) \right]       
\label{self-consistency}
\end{equation}
\noindent
where $f_{F}(\epsilon)=(e^{\epsilon/(k_{B}T)} +1)^{-1}$ is the Fermi function at temperature $T$ ($k_{B}$ being Boltzmann constant) and 
$v_{0}$ is the bare coupling constant of the contact interaction.
A suitable regularization of the inhomogeneous gap equation (\ref{self-consistency}) was implemented in Appendix B of Ref.~\cite{SPS-2013}, 
by drawing elements from the derivation of the Gross-Pitaevskii equation \cite{Pitaevskii_Stringari-2003} for composite bosons that form in the BEC limit, which was obtained in Ref.~\cite{Pieri-2003} starting from the BdG equations.   
In the process, the bare coupling constant $v_{0}$ that enters Eq.~(\ref{self-consistency}) gets replaced by the (dimensionless) coupling parameter 
$(k_{F} a_{F})^{-1}$, where $k_{F}=(3 \pi^{2} n_{0})^{1/3}$ is the Fermi wave vector associated with the (bulk) number density $n_{0}$ and $a_{F}$ is the scattering length of the two-fermion problem.
This coupling parameter enables one to span the whole BCS-BEC crossover \cite{Physics-Reports-2018}.
In practice, the crossover between the BCS and BEC regimes is exhausted within the range $-1 \lesssim (k_{F} a_{F})^{-1} \lesssim +1$ about the unitary limit (UL) where $(k_{F}a_{F})^{-1} = 0$.

When looking for bound states inside a vortex like in the present context, coupling values on the BCS side of unitarity should primarily be considered, since it will turn out that the number of bound states rapidly vanishes upon entering the BEC side of unitarity.
In addition, on the BEC side of the crossover at finite temperature it would be necessary to include pairing fluctuations beyond mean field \cite{Physics-Reports-2018}, a task which is beyond the purposes of the present paper.
On the other hand, on the BCS side of the crossover we shall find it necessary to extend the calculations down to the (numerically rather demanding) coupling value $(k_{F} a_{F})^{-1} = -4.0$, in order to recover what would be expected on the basis of the standard BCS approach.

For an isolated vortex with unit circulation and directed along the $z$ axis, the spatially dependent gap parameter is written in cylindrical coordinates as 
$\Delta(\mathbf{r}) = \Delta(\rho,\varphi,z) = \Delta(\rho) e^{i \varphi}$ with $\Delta(\rho)$ real.
Correspondingly, the eigenfunctions of the BdG equations (\ref{BdG-equations}) take the form: 
\begin{eqnarray}
u_{\nu}(\mathbf{r}) & = & u_{\nu_{\mathrm{r}},\ell, k_{z}}(\rho) \, \frac{e^{i \ell \varphi}}{\sqrt{2 \pi}} \,\, \frac{e^{i k_{z} z}}{\sqrt{2 \pi}}
\label{explicit-form-u} \\
\!\! v_{\nu}(\mathbf{r}) & = & v_{\nu_{\mathrm{r}},\ell, k_{z}}(\rho) \, \frac{e^{i (\ell - 1) \varphi}}{\sqrt{2 \pi}} \,\, \frac{e^{i k_{z} z}}{\sqrt{2 \pi}}
\label{explicit-form-v}
\end{eqnarray}
where $\nu_{\mathrm{r}}$ is the radial quantum number, $\ell$ is an integer (both positive and negative), and $\{u_{\nu_{\mathrm{r}},\ell, k_{z}}(\rho),v_{\nu_{\mathrm{r}},\ell, k_{z}}(\rho)\}$ are real functions for the bound states of primary interest here.
Within this approach, for given values of $\ell$ and $k_{z}$ the BdG equations are numerically integrated outwards, starting from $\rho = 0$ with suitable indicial conditions, up to a maximum value $R_{\mathrm{out}}$ 
(values ranging from $k_{F} R_{\mathrm{out}} = 60$ for $(k_{F} a_{F})^{-1} = 0$ up to $k_{F} R_{\mathrm{out}} = 1500$ for $(k_{F} a_{F})^{-1} = -4.0$ proves sufficient for practical purposes).
At that point, ``free'' boundary conditions are enforced at $\rho = R_{\mathrm{out}}$ by using suitable linear combinations of Bessel, Neumann, and Hankel functions, in terms of which the solutions of the BdG equations can be expressed for $\rho \ge R_{\mathrm{out}}$.
This procedure offers a definite advantage, in that it avoids the common practice of constraining the vortex in a cylinder with infinite walls at $\rho = R_{\mathrm{out}}$.
For this reason, this procedure allows one to clearly distinguish truly bound states (which decay exponentially 
for $\rho \gg R_{\mathrm{out}}$) from continuum states (with oscillating behavior extending up to infinity).
In practice, we have considered values of $\ell$ not smaller than $200$ and values of $|k_{z}|$ up to $3 k_{F}$.

In this way, for given $k_{z}$ and irrespective of the value of $\ell$, a first \emph{energy threshold} between bound and continuum states occurs at the (temperature-dependent) value $\Delta_{0}$ recovered by 
$\Delta(\rho)$ deep in the bulk region when $\tilde{\mu} >0$ (where $\tilde{\mu}  = \mu - k_{z}^{2}/2m$), while a second threshold occurs at $\sqrt{\Delta_{0}^{2} + \tilde{\mu}^{2}}$ when $\tilde{\mu} <0$.
Accordingly, for $0<\varepsilon<\Delta_{0}$ only bound states can be found, for $\Delta_{0}<\varepsilon<\sqrt{\Delta_{0}^{2} + \tilde{\mu}^{2}}$ bound states embedded in the continuum (with different values 
of $k_{z}$) can also be found, and for $\sqrt{\Delta_{0}^{2} + \tilde{\mu}^{2}}<\varepsilon$ only continuum states occur.

Once the eigenfunctions $\{u_{\nu_{\mathrm{r}},\ell, k_{z}}(\rho),v_{\nu_{\mathrm{r}},\ell, k_{z}}(\rho)\}$ have been determined together with the corresponding eigenvalues $\left\{\varepsilon_{\nu_{\mathrm{r}},\ell, k_{z}}\right\}$ according to the above procedure, one can obtain the  \emph{local density of states} given by:
\begin{eqnarray}
N(\rho;E) & = & \int_{-\infty}^{+\infty} \! \frac{d k_{z}}{(2 \pi)^{2}} \sum_{\nu_{\mathrm{r}},\ell} 
\left[ |u_{\nu_{\mathrm{r}},\ell, k_{z}}(\rho)|^{2} \, \delta(E - \varepsilon_{\nu_{\mathrm{r}},\ell, k_{z}}) \right.
\nonumber \\
& + & \left . |v_{\nu_{\mathrm{r}},\ell, k_{z}}(\rho)|^{2} \, \delta(E + \varepsilon_{\nu_{\mathrm{r}},\ell, k_{z}}) \right] 
\label{local-density_of_states}
\end{eqnarray}
\noindent
which has dimensions of [volume$\times$energy]$^{-1}$.
This quantity, which is of primary experimental interest \cite{RMP-2007,Esteve-2008}, contains contributions from both bound and continuum states \cite{footnote}.
We are now in a position to calculate $N(\rho;E)$ as a function of coupling along the BCS-BEC crossover and of temperature in the superfluid phase. 
In addition, by our approach, we can also distinguish the separate contributions to $N(\rho;E)$ from bound and continuum states.
 
 It is further relevant to point out that, contrary to a common assumption that the radial vortex profile approaches asymptotically the bulk value 
 $\Delta_{0}$ in an exponential way \cite{Berthod-2016,Tempere-2017}, the gap parameter $\Delta(\rho)$ has instead \emph{a long tail with
 a power-law dependence} of the type $\Delta_{0} \left( 1 - \frac{\zeta^{2}}{2 \rho^{2}} \right)$.
This property, which went apparently unnoticed in the literature, was found to be valid for all couplings throughout the BCS-BEC crossover by the detailed numerical analysis carried out in Ref.~\cite{SPS-2013}. 
As shown in Appendix \ref{sec:appendix}, this property can be also captured analytically in terms of the Gross-Pitaevskii equation \cite{Pitaevskii_Stringari-2003}, to which the BdG equations have been shown to reduce in the BEC (strong-coupling) limit \cite{Pieri-2003}.
It is actually in the more limited radial range $k_{F}^{-1} \le \rho \lesssim 5 R_{\mathrm{v}}$ only that the vortex profile has an exponential dependence, from which a characteristic (temperature and coupling dependent) coherence length $\xi$ can be determined, where
$R_{\mathrm{v}}$ marks the position of the maximum value of the radial current \cite{SPS-2013}.
As is turns out from the numerical analysis of Ref.~\cite{SPS-2013} (and confirmed by from the present one), the two length scales $\zeta$ and $\xi$ coincide with each other within numerical error for all couplings and temperatures.

Accordingly, in the following we shall find it convenient to partition the vortex profile into an ``inner'' region (where the exponential decay applies) and an
``outer'' region (where the power-law behavior takes over), the boundary between the two regions being (approximately) taken at $\rho \approx \xi$.
The counting of the number of bound states will similarly be partitioned.
We are then going to distinguish whether a given bound state with quantum numbers ($\nu_{\mathrm{r}},\ell,k_{z}$) belongs to the inner or to the outer region, by looking at the (dimensionless) partial normalization condition:
\begin{equation}
\mathcal{P}_{\nu_{\mathrm{r}},\ell, k_{z}}(\xi) = \int_{0}^{\xi} \! d\rho \, \rho \, \left[ u_{\nu_{\mathrm{r}},\ell,k_{z}}(\rho)^{2} \, + \, v_{\nu_{\mathrm{r}},\ell, k_{z}}(\rho)^{2}  \right] \, .
\label{partial-normalization-bound-states}
\end{equation}
\begin{figure*}[t]
\begin{center}
\includegraphics[width=15.0cm,angle=0]{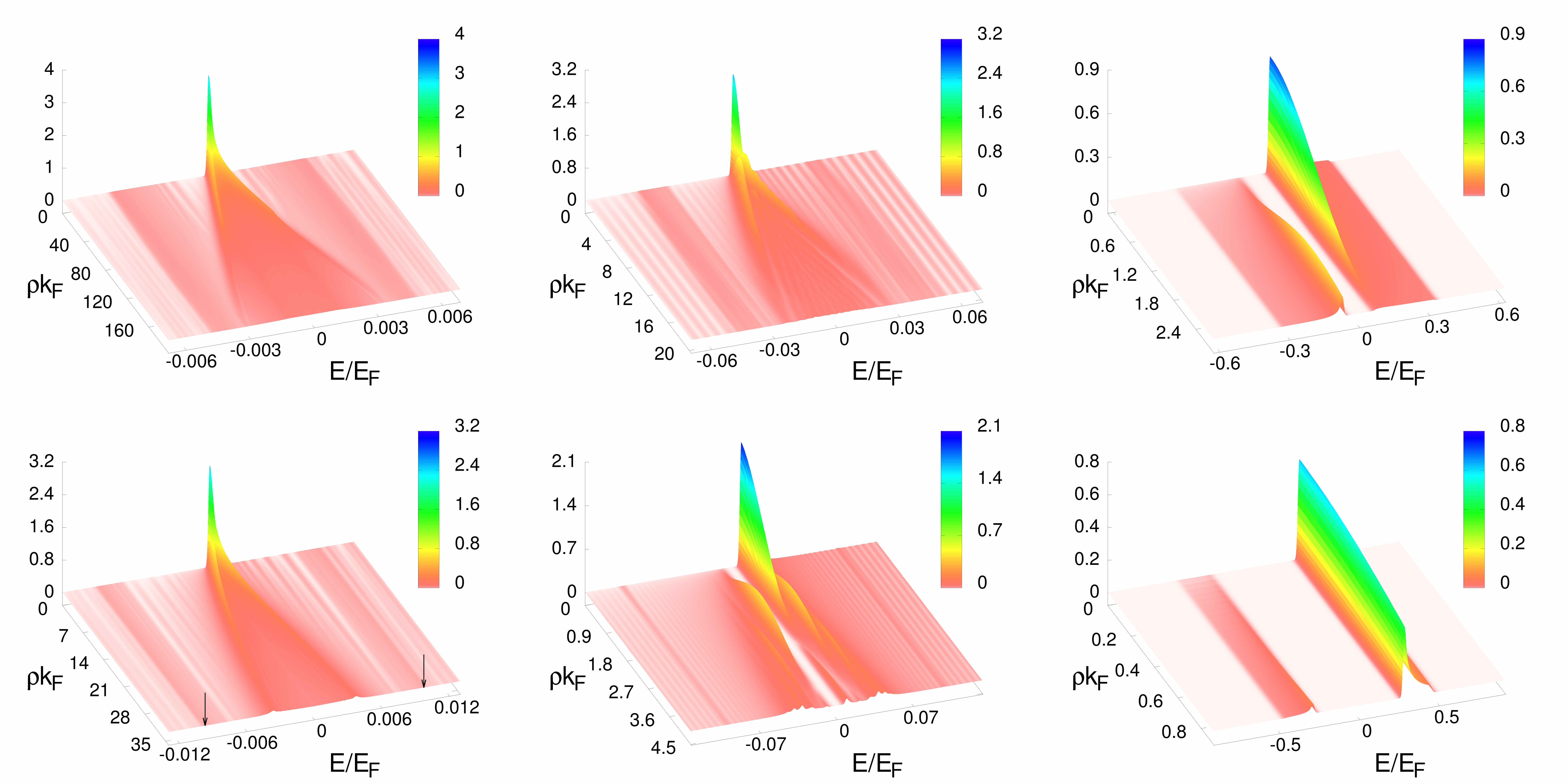}
\caption{(Color online) The local density of states $N(\rho;E)_{\mathrm{ib}}$, contributed by the \emph{inner bound} states, 
                                     is shown for three couplings $(k_{F}a_{F})^{-1}=(-3.0,-1.5,0.0)$ (from left to right) and two temperatures $T=(0,0.9)T_{c}$ (from bottom to top).
                                     Here, $N(\rho;E)_{\mathrm{ib}}$ is in units of $k_{F}^{3}/E_{F}$, $\rho$ in units of $k_{F}^{-1}$, and $E$ in units of  $E_{F}$ (where $E_{F}=k_{F}^{2}/2m$ is the Fermi energy).
                                     In the left-bottom panel, the arrows pointing toward the energy scale mark the values $\pm \Delta_{0}$ of the continuum threshold at the corresponding coupling and temperature
                                     (cf. Table~\ref{Table-I} below).}
\label{Figure-1}
\end{center}
\end{figure*} 
We then attribute the location of the bound state to the inner (outer) region when $\mathcal{P}_{\nu_{\mathrm{r}},\ell, k_{z}}(\xi)$ exceeds (falls behind), say, $0.5$ as compared 
to the value $1.0$ of the complete normalization when the upper limit of the integral in Eq.~(\ref{partial-normalization-bound-states}) 
extends to infinity.
Owing to the ($\rho^{-2}$) power-law tail of the gap parameter, it turns out that the vortices residing in the outer region by far exceed in number
those residing in the inner region, although most of them lie far away from the vortex center.

In practice, the bound states in the inner region (and therefore their number) are numerically much more under control than those in the outer region, especially when the latter ones hinge mostly close to the boundary at $R_{\mathrm{out}}$ where the connection with decaying exponentials is enforced through the boundary conditions.
In addition, when several vortices would be packed up together to form a vortex lattice, the nature of the bound states in the remote external part of the outer region of what would have been a vortex in isolation is expected to be strongly modified by the presence of the surrounding vortices.
For this reason, it may anyway not be worth to insist in extending the size of the outer region beyond those values of $R_{\mathrm{out}}$ that we have considered in practice for the numerical calculations.

\section{Numerical results} 
\label{sec:numerical_results}

We pass to characterize the presence of bound states in an isolated vortex, for all temperatures in the superfluid phase and couplings along the BCS-BEC crossover, by taking advantage of our accurate numerical calculations in terms of the BdG equations following the methodologies developed in Ref.~\cite{SPS-2013}.
We shall find that the number of bound states rapidly decreases on the BEC side of unitarity, so that, in practice, only the coupling region from the BCS regime up to unitarity will be relevant to our purposes.
Several features will be highlighted by looking at the details of the local density of states, or simply by counting the number of bound states that can be identified for given coupling and temperature.
Our main findings are summarized by the following features.

\begin{figure*}[t]
\begin{center}
\includegraphics[width=15.0cm,angle=0]{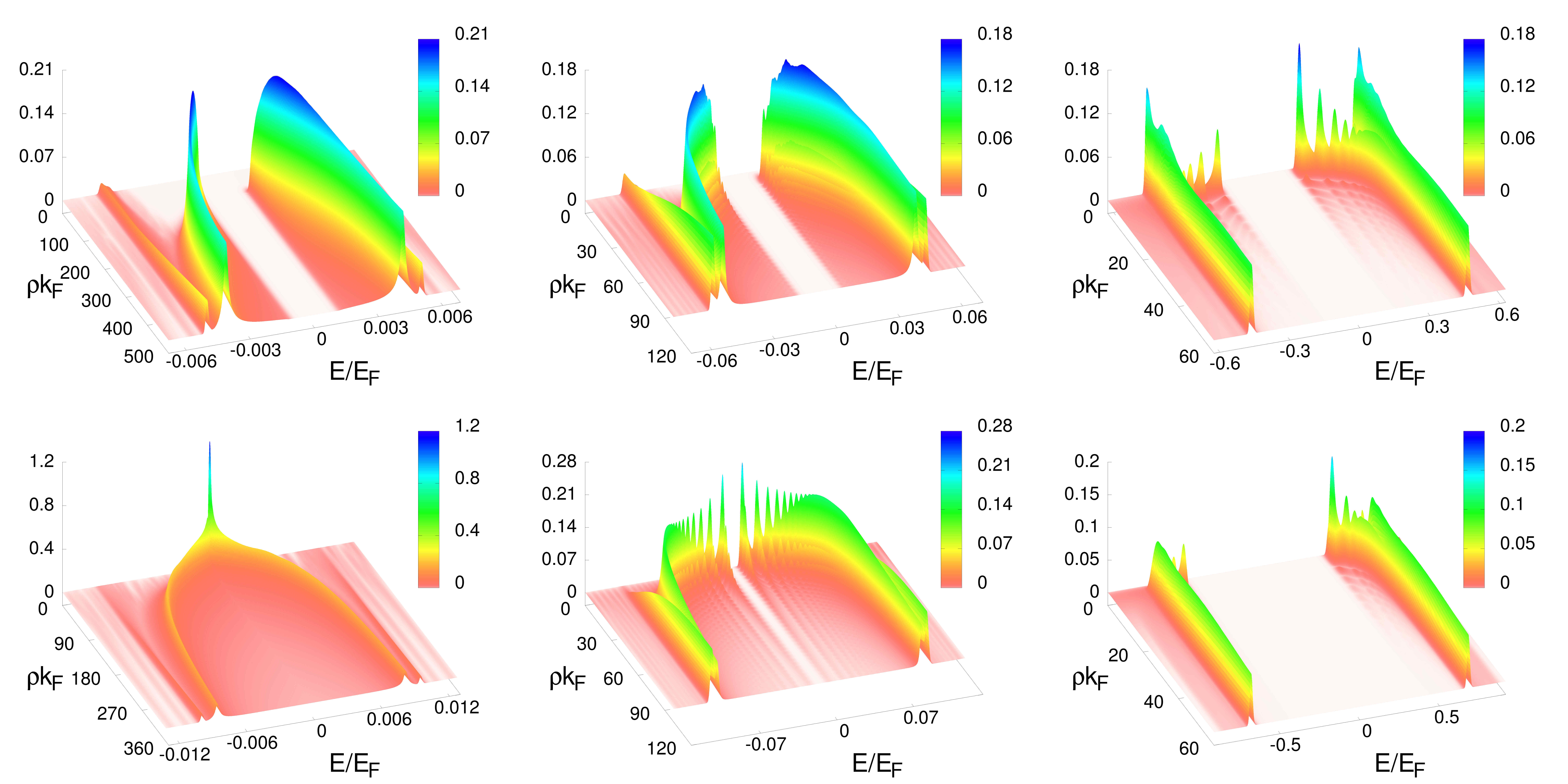}
\caption{(Color online) The local density of states $N(\rho;E)_{\mathrm{ob}}$, contributed by the \emph{outer bound} states, 
                                     is shown for three couplings $(k_{F}a_{F})^{-1}=(-3.0,-1.5,0.0)$ (from left to right) and two temperatures $T=(0,0.9)T_{c}$ (from bottom to top).
                                     Like in Fig.~\ref{Figure-1}, $N(\rho;E)_{\mathrm{ob}}$ is in units of $k_{F}^{3}/E_{F}$, $\rho$ in units of $k_{F}^{-1}$, and $E$ in units of  $E_{F}$.}
\label{Figure-2}
\end{center}
\end{figure*} 

\vspace{0.05cm}
\begin{center}
{\bf A. Contribution from bound and scattering states to the local density of states}
\end{center}

One of the main questions to be asked at the outset is where the various contributions, in which the local density of states $N(\rho;E)$ of Eq.~(\ref{local-density_of_states}) can be partitioned, are spatially located over the extension of an isolated vortex.
These contributions can be identified as those corresponding to the bound states residing essentially in the inner and outer regions of the vortex, and as that corresponding to the scattering states.
Here, the partial normalization criterion (\ref{partial-normalization-bound-states}) is used to assign a given bound state either to the inner or outer regions of the vortex (see also subsection~\ref{sec:numerical_results}-B below).  

\begin{figure*}[t]
\begin{center}
\includegraphics[width=15.0cm,angle=0]{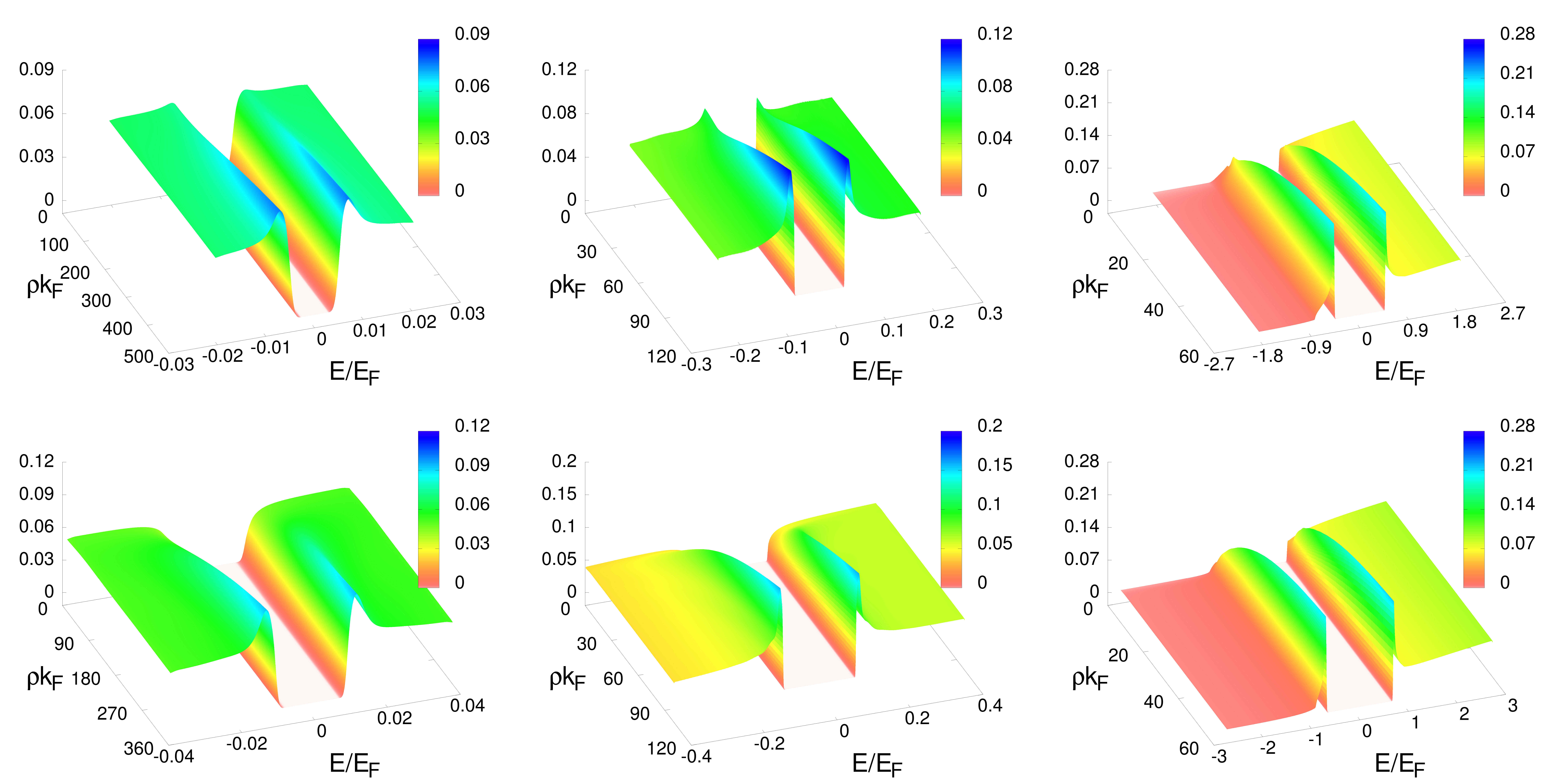}
\caption{(Color online) The local density of states $N(\rho;E)_{\mathrm{sc}}$, contributed by the \emph{scattering states}, 
                                     is shown for three couplings $(k_{F}a_{F})^{-1}=(-3.0,-1.5,0.0)$ (from left to right) and two temperatures $T=(0,0.9)T_{c}$ (from bottom to top).
                                     Like in Figs.~\ref{Figure-1} and \ref{Figure-2}, $N(\rho;E)_{\mathrm{sc}}$ is in units of $k_{F}^{3}/E_{F}$, $\rho$ in units of $k_{F}^{-1}$, 
                                     and $E$ in units of  $E_{F}$.
                                     The white stripe around $E=0$ corresponds to the gap region where no scattering states can occur and the associated local density of states identically 
                                     vanishes.}
\label{Figure-3}
\end{center}
\end{figure*}

Accordingly, we express the local density of states $N(\rho;E)$ of Eq.~(\ref{local-density_of_states}) as the sum of three contributions, namely, 
$N(\rho;E)_{\mathrm{ib}}$, $N(\rho;E)_{\mathrm{ob}}$, and $N(\rho;E)_{\mathrm{sc}}$, where ``ib'' stands for \emph{inner-bound}, ``ob'' for \emph{outer-bound}, and ``sc'' for \emph{scattering}.
These three contributions are reported in Figs.~\ref{Figure-1}, \ref{Figure-2}, and \ref{Figure-3}, respectively, as a function of both $\rho$ and $E$, in each case for three characteristic 
couplings $(k_{F}a_{F})^{-1}=(-3.0,-1.5,0.0)$ and two significant temperatures $T=(0,0.9)T_{c}$.
[One should, however, be aware that, due to the progressive importance of pairing fluctuations beyond mean field at finite temperature for increasing coupling, the data at unitarity 
and finite temperature are expected to be quantitatively less reliable than those at weaker couplings.]

\begin{table}[h]
\begin{tabular}{ cccc } 
\hline
\hline
$(k_{F} a_{F})^{-1}$ & $T=0$  & $T=0.9 \, T_{c}$  &  $T_{c}/E_{F}$ \\
\hline
 0.0  & 0.6864 & 0.3445  & 0.4965  \\      
 -1.5 & 0.1001 & 0.0523  & 0.0579 \\ 
  -3.0 & 0.0097 & 0.0051 & 0.0055 \\ 
\hline
\hline
\end{tabular}
\caption{Values of $\Delta_{0}(T)$ (in units of $E_{F}$) for the couplings and temperatures $T=(0,0.9)T_{c}$ considered in the panels of Figs.~\ref{Figure-1}-\ref{Figure-3}.
The last column gives the corresponding values of the critical temperature $T_{c}$ (in units of $E_{F}$).}
\label{Table-I}
\end{table}
 
From Figs.~\ref{Figure-1}-\ref{Figure-3} one notices that the contribution $N(\rho;E)_{\mathrm{ib}}$ is mostly spatially concentrated over the inner portion of the vortex, while the contribution $N(\rho;E)_{\mathrm{ob}}$ is spread over a much wider spatial region.
On the other hand, $N(\rho;E)_{\mathrm{sc}}$ at low energies above (the temperature dependent) threshold is depressed in the central region of the vortex owing to the orthogonality requirement of quantum mechanical states with different energies, while at high energies above threshold $N(\rho;E)_{\mathrm{sc}}$ is almost uniformly spread over all space owing to the occurrence of spatial oscillations with small wavelength.
The situation remains qualitatively the same also for increasing temperature from $T= 0$ to $T=0.9T_{c}$.
More marked differences appear instead for increasing coupling, from $(k_{F}a_{F})^{-1}=-3.0$ deep in the weak-coupling (BCS) region to unitarity where $(k_{F}a_{F})^{-1}=0$, along which one notices
a progressively enhanced asymmetry between negative and positive energies.
We shall return below to a more detailed discussion of this asymmetry.

From Figs.~\ref{Figure-1}-\ref{Figure-3} one also notices that, for each coupling and temperature, a sharp threshold for the continuum states occurs at $E=\Delta_{0}(T)$ 
(cf. Fig.~\ref{Figure-3}), and that bound states ``embedded'' in the continuum are also found when $E>\Delta_{0}$ (cf. Figs.~\ref{Figure-1}-\ref{Figure-2}), although they give 
a small overall contribution to $N(\rho;E)$.
This can be seen, in particular, in the left-bottom panel of Fig.~\ref{Figure-1}, where the two arrows mark the values $\pm \Delta_{0}$ of the continuum threshold beyond which bound states embedded in the continuum appear.
For convenience, the numerical values of $\Delta_{0}(T)$ relevant to Figs.~\ref{Figure-1}-\ref{Figure-3} are reported in Table~\ref{Table-I}.

We have further verified that, past unitarity upon approaching the BEC regime, there occurs a progressive depletion of $N(\rho;E)_{\mathrm{ib}}$ in the central region of the vortex, implying that bound states tend to disappear in this coupling regime.
In addition, for given coupling this depletion is essentially complete for $T=0$, while it becomes only partial upon increasing the temperature.
We shall return below to a more detailed discussion of this feature.
 
\vspace{0.05cm}
\begin{center}
{\bf B. Counting the number of bound states in the inner and outer regions}
\end{center}

\begin{figure}[h]
\begin{center}
\includegraphics[width=8.0cm,angle=0]{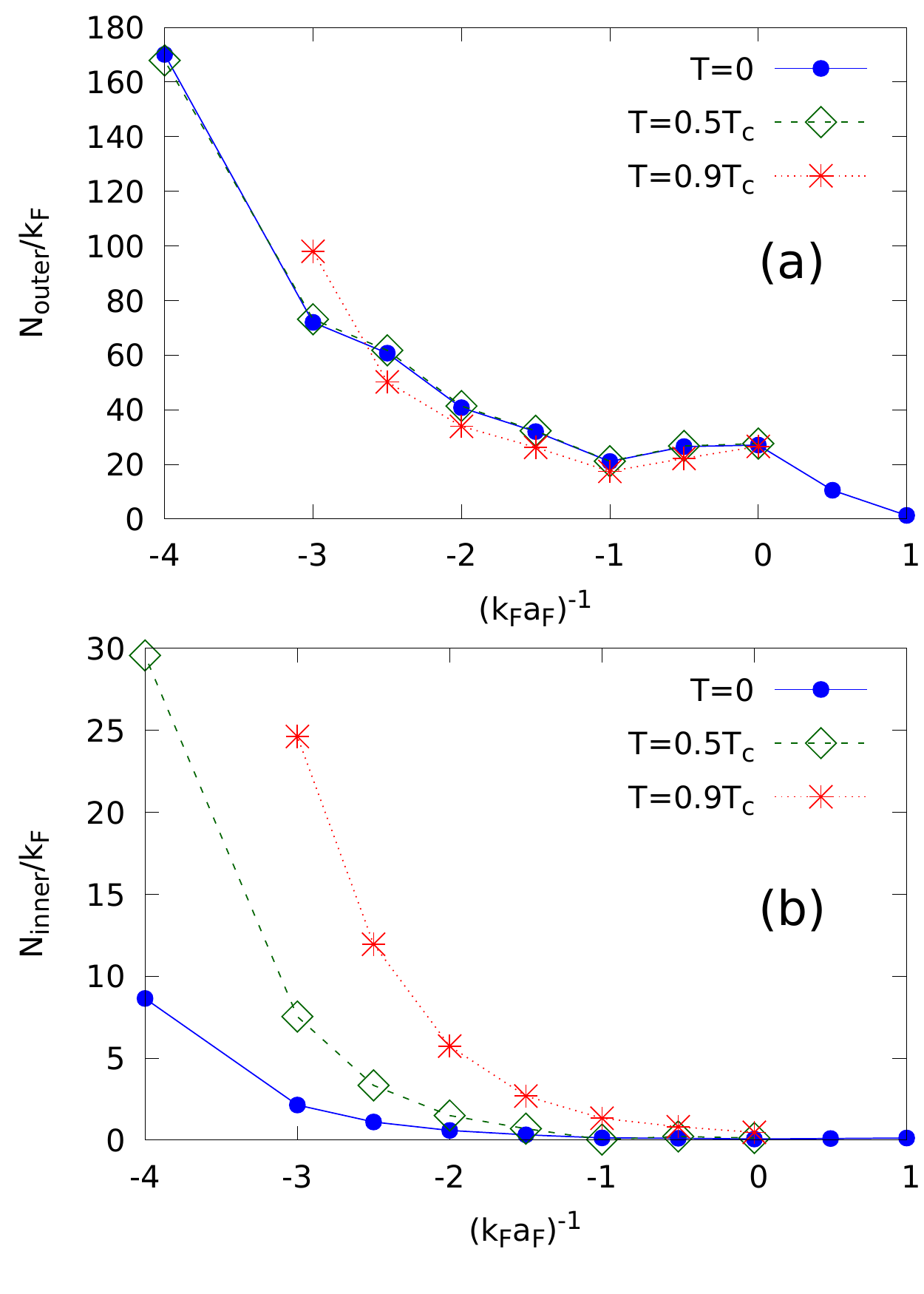}
\caption{(Color online) The line density of bound states $N_{\mathrm{inner}}$ in the inner region (lower panel) and $N_{\mathrm{outer}}$ in the outer region (upper panel) 
                                    of the vortex vs the coupling $(k_{F}a_{F})^{-1}$ for three temperatures: $T=0$ (filled circles), $T=0.5T_{c}$ (open boxes), 
                                    and $T=0.9T_{c}$ (stars).
                                    Both $N_{\mathrm{inner}}$ and $N_{\mathrm{outer}}$ have been divided by $k_{F}$ to make them dimensionless.}
\label{Figure-4}
\end{center}
\end{figure} 

Another relevant piece of information can be obtained by a simple count of the number of bound states, separately in the inner and outer regions as defined above.
To this end, we recall that the bound-state wave functions are localized in the $x$-$y$ plane orthogonal to the $z$ axis of the vortex, so that a whole branch spanned by the wave vector $k_{z}$ is associated with each bound state with given angular momentum $\ell$ and radial quantum number $\nu_{\mathrm{r}}$.
Accordingly, in the following the numbers of bound states will be given per unit length along the direction of the vortex axis, since they will be summed (integrated) over the wave vector $k_{z}$ 
(as it is also done in Eq.~(\ref{local-density_of_states})).
With this procedure, the line density of bound states $N_{\mathrm{inner}}$ and $N_{\mathrm{outer}}$, that reside, respectively, in the inner and outer regions of an isolated vortex, can be obtained 
as a function of coupling and temperature.

Figure~\ref{Figure-4} shows both $N_{\mathrm{inner}}$ (lower panel) and $N_{\mathrm{outer}}$ (upper panel) for three characteristic temperatures $T=(0,0.5,0.9)T_{c}$, with the coupling $(k_{F} a_{F})^{-1}$ ranging from $-4.0$ deep in the BCS regime up to unitarity.
In addition, for $T=0$ the calculation has been extended to $(k_{F} a_{F})^{-1} = 1.0$ on the BEC side of unitarity.

A few general features can be evidenced from these plots:
(i) Both $N_{\mathrm{inner}}$ and $N_{\mathrm{outer}}$ rapidly decrease from the BCS regime to the UL and quickly vanish once the BEC side past unitarity is reached, thus implying that the bound states have no relevance for the composite bosons that form out of fermion pairs in the BEC regime;
(ii) For given coupling, $N_{\mathrm{inner}}$ increases upon increasing the temperature, whereby the depth $\Delta_{0}$ of the vortex decreases but at the same time its width $\xi$ increases;
(iii) For given coupling and temperature, $N_{\mathrm{outer}}$ by far exceeds $N_{\mathrm{inner}}$, although $N_{\mathrm{outer}}$ appears to be essentially independent of temperature.

It should be remarked that the values of $N_{\mathrm{outer}}$ we have obtained are numerically less reliable that those of $N_{\mathrm{inner}}$, because they depend on the value of the radius $R_{\mathrm{out}}$ where the boundary conditions are enforced on the solutions of the BdG equations (cf. Section~\ref{sec:formal_aspects}).
In this context, we found it numerically too demanding to make this radius exceed $k_{F} R_{\mathrm{out}} = 1500$ for the smaller coupling we could reach deep in the BCS regime.
For these reasons, in the following we shall limit ourselves to further examine the dependence of $N_{\mathrm{inner}}$ on $\Delta_{0}$ and $\xi$, leaving aside similar considerations for $N_{\mathrm{outer}}$.	

\begin{figure}[t]
\begin{center}
\includegraphics[width=8.0cm,angle=0]{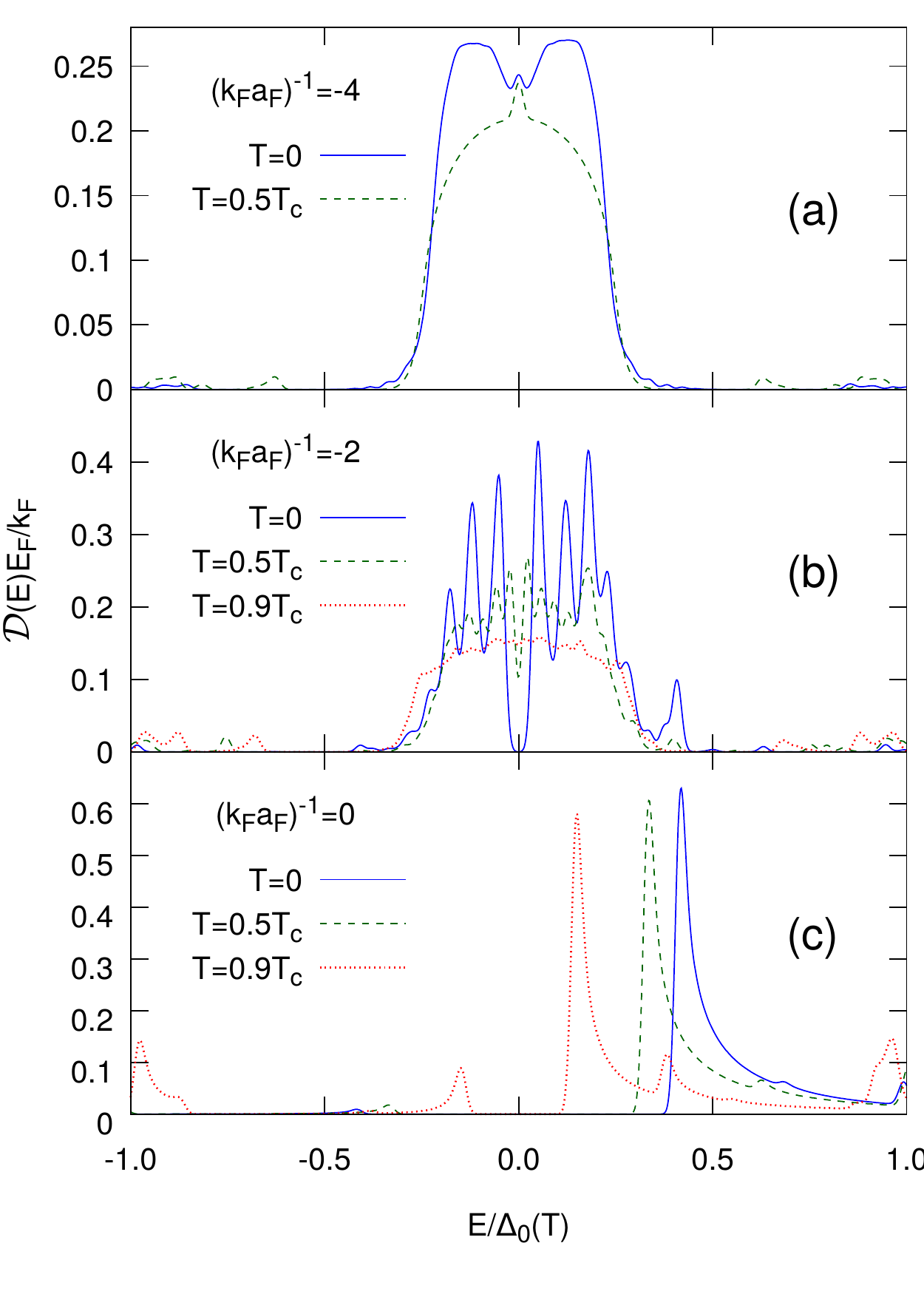}
\caption{(Color online) The quantity $\mathcal{D}(E)$ of Eq.~(\ref{local-DOF-near-vortex-center}) (in units of $k_{F}/E_{F}$) 
                                     is shown vs $E$ (in units of the bulk gap $\Delta_{0}(T)$ at temperature $T$), for the couplings $(k_{F}a_{F})^{-1}=-4.0$ (upper panel), 
                                     $(k_{F}a_{F})^{-1}=-2.0$ (middle panel), and $(k_{F}a_{F})^{-1}=0.0$ (lower panel), and for the temperatures $T=0$ (full line), 
                                     $T=0.5T_{c}$ (dashed line), and $T=0.9T_{c}$ (dotted line).}
\label{Figure-5}
\end{center}
\end{figure} 

\vspace{0.05cm}
\begin{center}
{\bf C. Symmetry vs asymmetry of the local density of states about the vortex center}
\end{center}

A characteristic feature, that distinguishes whether a homogeneous attractive Fermi gas lies in the weak-coupling (BCS) or the unitary regime, is the \emph{degree of asymmetry} of the single-particle spectral function, with the two peaks of this spectral function being much more symmetric about zero frequency in the BCS regime than at the UL \cite{PPPS-2012}.
The occurrence of particle-hole asymmetry is a characteristic feature of the BCS-BEC crossover, whereby the underlying Fermi surface gets progressively modified when passing 
from the BCS to the BEC limits across UL, as the ratio $\Delta_{0}/E_{F}$ increases.
Our interest here is to verify to what extent this characteristic feature is maintained also for the present inhomogeneous problem, such that a measurement of the local density of states 
about the vortex center would be able to detect whether or not the superfluid properties of the Fermi gas deviate from those of an ordinary BCS superconductor.

As we have already noted while commenting Figs.~\ref{Figure-1}-\ref{Figure-3}, all three contributions to the local density of states $N(\rho;E)$ that we have considered become progressively more asymmetric when passing from the BCS to the unitary regime.
A more quantitative characterization of this asymmetry can be obtained as follows.
Let us consider a spatial region in the inner part of the vortex, say, of linear extension $\xi/2$ around the vortex center at $\rho = 0$
(in such a way to concentrate our attention close to the center of the vortex).
We then look for the behavior of the integrated quantity 
\begin{equation}
\mathcal{D}(E) = 2 \pi \! \int_{0}^{\xi/2} \!\!\! d\rho \, \rho \, N(\rho;E) 
\label{local-DOF-near-vortex-center}
\end{equation}
where $N(\rho;E)$ is the local density of states (\ref{local-density_of_states}).
The quantity (\ref{local-DOF-near-vortex-center}) has dimensions of [length$\times$energy]$^{-1}$ and contains, in principle, contributions from both bound and continuum states, 
depending on the value of $E$.
In the following, we shall be interested in the energy interval $-\Delta_{0}(T) \le E \le \Delta_{0}(T)$ where $\Delta_{0}(T)$ is the bulk value of the gap parameter at temperature $T$, such that
only bound states contribute to $\mathcal{D}(E)$ in this energy interval.

\begin{figure}[t]
\begin{center}
\includegraphics[width=8.6cm,angle=0]{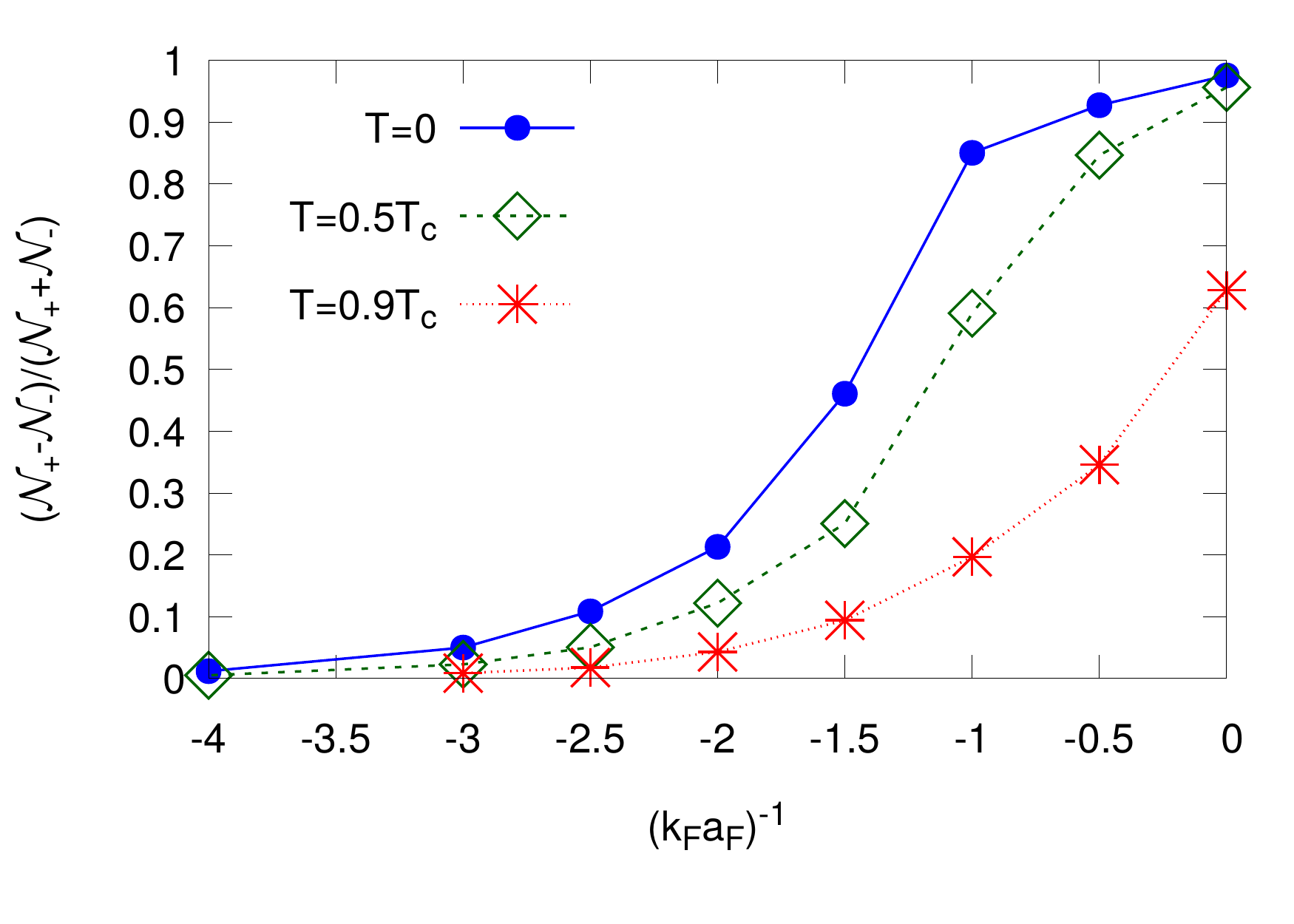}
\caption{(Color online) Asymmetry parameter, defined in terms of the quantities (\ref{N-plus}) and (\ref{N-minus}), as a function of the coupling $(k_{F} a_{F})^{-1}$ for three different
                                     temperatures: $T=0$ (filled circles), $T=0.5T_{c}$ (open boxes), and $T=0.9T_{c}$ (stars).}
\label{Figure-6}
\end{center}
\end{figure} 

Plots of $\mathcal{D}(E)$ vs $E$ are shown in Fig.~5 for three characteristic couplings $(k_{F}a_{F})^{-1}=(-4.0,-2.0,0.0)$ and three characteristic temperatures $T=(0,0.5,0.9)T_{c}$ for each coupling (with the exception of the coupling $(k_{F}a_{F})^{-1}=-4.0$ for which $T=0.9T_{c}$ is lacking).
Irrespective of temperature, drastic changes are seen to occur in the shape of $\mathcal{D}(E)$, which becomes progressively more asymmetric when passing from $(k_{F}a_{F})^{-1}=-4.0$ (upper panel) to 
$(k_{F}a_{F})^{-1}=0.0$ (lower panel).

An even sharper characterization, about the development of this asymmetry when approaching the UL from the BCS side, can be obtained by integrating $\mathcal{D}(E)$ over either the positive or negative portions of the gap region, thereby defining  
\begin{eqnarray}
\mathcal{N}_{+} & = & \int_{0}^{\Delta_{0}(T)} \!\!\! dE \,\, \mathcal{D}(E)
\label{N-plus} \\
\mathcal{N}_{-} & = & \int_{-\Delta_{0}(T)}^{0} \! \!\!dE \,\, \mathcal{D}(E)
\label{N-minus}
\end{eqnarray}
which have dimensions of [length]$^{-1}$.
In this way, a suitable \emph{asymmetry parameter} $(\mathcal{N}_{+} - \mathcal{N}_{-})/(\mathcal{N}_{+} + \mathcal{N}_{-})$ can then be introduced, which can be analyzed as a function of coupling and temperature.
The results are reported in Fig.~\ref{Figure-6}.
One sees that, essentially for all temperatures, there occurs a steep increase of this asymmetry parameter, from the value zero in the BCS regime to almost unity at UL, which at $T=0$ has reached half the way at about the coupling $(k_{F} a_{F})^{-1}=-1.5$.
This quantity thus provides one with a quick clue for the closeness of the superfluid Fermi gas to the UL, by looking at the distribution of the bound states about the vortex center.

From the experimental results available from Ref.~\cite{Berthod-2017}, no clear evidence for the occurrence of particle-hole asymmetry can apparently be extracted. 
One has to consider, however, that in Ref.~\cite{Berthod-2017} the presence of bound states in vortices was evidenced in a high-$T_{c}$ cuprate superconductor for the first time, through a delicate subtraction procedure of a significant background. Additional detailed measurements of density-of-states spectra would then be required to evidence a particle-hole asymmetry 
over and above this subtraction.

\vspace{0.05cm}
\begin{center}
{\bf D. A scaling relation}
\end{center}

\begin{figure}[t]
\begin{center}
\includegraphics[width=9.3cm,angle=0]{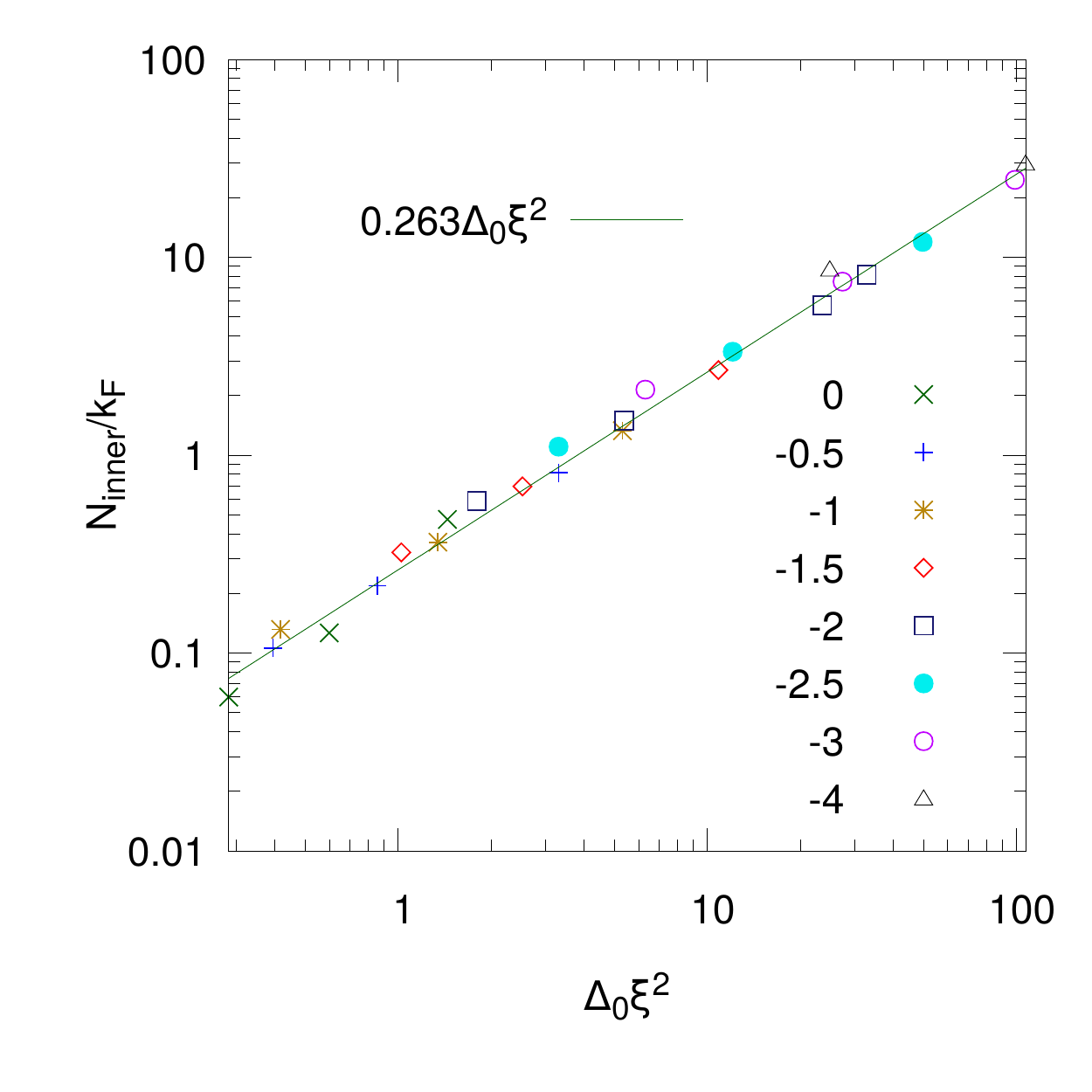}
\caption{(Color online) Scaling plot showing the values of the line density of bound states $N_{\mathrm{inner}}$ (in units of $k_{F}$), obtained for given coupling and temperature, reported 
                                    as a function of the variable $\Delta_{0} \xi^{2}$, where the values of $\Delta_{0}$ and $\xi$ are obtained at given coupling and temperature.
                                    For each coupling value listed in the figure, the values of $N_{\mathrm{inner}}$ correspond to the three temperatures $T=(0,0.5,0.9)T_{c}$
                                    (with the exception of the coupling $(k_{F}a_{F})^{-1}=-2.0$ for which the value for $T=0.95 \, T_{c}$ is also added, 
                                    and of the coupling $(k_{F}a_{F})^{-1}=-4.0$ for which the value for $T=0.9 \, T_{c}$ is missing).}
\label{Figure-7}
\end{center}
\end{figure} 

In Fig.~\ref{Figure-4}, $N_{\mathrm{inner}}$ was found to depend separately on coupling and temperature.
However, both the depth $\Delta_{0}$ (or bulk value of the gap parameter) and the width $\xi$ of the vortex also depend on coupling and temperature.
It will then be interesting to eliminate this double dependence and express $N_{\mathrm{inner}}$ directly as a function of a \emph{single} variable obtained by combining $\Delta_{0}$ and $\xi$, 
irrespective of the corresponding values of coupling and temperature.
Through this attempt, we have found that the product $\Delta_{0} \xi^{2}$ is the appropriate \emph{scaling variable} on which the line density of bound states $N_{\mathrm{inner}}$ effectively depends.

Accordingly, we have organized the numerical values $N_{\mathrm{inner}}$ from Fig.~\ref{Figure-4} into the single plot shown in Fig.~\ref{Figure-7}, where
$N_{\mathrm{inner}}$ is reported vs $\Delta_{0} \xi^{2}$ over a double-log scale so as to put on equal footing quite different sets of data (spanning four decades, from $10^{-2}$ to $10^{2}$). 
Here, $\xi$ is is units of the inverse of the Fermi wave vector $k_{F}$ and $\Delta_{0}$ in units of the Fermi energy $E_{F}$, such that the variable $\Delta_{0} \xi^{2}$ of 
Fig.~\ref{Figure-7} is dimensionless.

The resulting linear scaling dependence of $N_{\mathrm{inner}}$ vs $\Delta_{0} \xi^{2}$ (which is evidenced in the figure by the straight line $N_{\mathrm{inner}}/k_{F} = 0.263 \, \Delta_{0} \xi^{2}$) 
appears quite remarkable, to the extent that essentially \emph{all} data with individually quite different values of $\Delta_{0}$ and $\xi$ fall on this straight line.
We have further verified that this scaling relation between $N_{\mathrm{inner}}$ and $\Delta_{0} \xi^{2}$ holds also when the counting of bound states residing in the inner region is enlarged, by shifting $\xi$, e.g., 
to $2 \xi$ in Eq.~(\ref{partial-normalization-bound-states}), provided that the same replacement $\xi \rightarrow 2 \xi$ is also made in the variable $\Delta_{0} \xi^{2}$ on which $N_{\mathrm{inner}}$ depends.

\section{Concluding remarks}
\label{sec:conclusions}

In this paper, we have given a detailed account about the number, energy location, spatial location, and shape of the large number of bound states that are present within an isolated vortex embedded in an otherwise homogeneous superfluid.
We have done this by an accurate solution of the BdG equations based on the methods developed in Ref.~\cite{SPS-2013}, for a dilute Fermi gas spanning the BCS-BEC crossover with the temperature ranging from zero to the superfluid critical temperature.
In this way, we have been able to provide a criterion for locating where a given Fermi gas lies along this crossover, in terms of the asymmetry between positive and negative energies
of the local density of states. 
From our numerical calculations we were also able to extract a universal scaling relation, that relates the number of bound states in the inner region of the vortex with the depth and width of the vortex itself.

We have further verified that bound states occur in the interior of a vortex for couplings on the BCS side of unitary, and rapidly disappear when entering the BEC regime past unitarity.
Consistently with this theoretical finding, a clear experimental finding for the occurrence of bound states in a vortex in a superconducting material \cite{Berthod-2017} can indeed be greeted as a signature that the superfluid phase of that material may be described by the conventional BCS theory \cite{SX-2017}.
Yet, it is also known from the analysis of Ref.~\cite{SPS-2013} that it is rather the continuum part of the spectrum of a vortex (as obtained by solving the BdG equations) that exhausts, in practice, most part of the contribution to physical quantities, such as the profile of the gap parameter and the local number density.

In this paper, the analysis has been limited to considering an $s$-wave order parameter, and the question may arise about the role of the symmetry of the order parameter on our results. 
In this context, we may refer to the results of Ref.~\cite{Kato-2001}, where a BdG calculation for an isolated vortex with a $d$-wave order parameter was reported, with the conclusion that the spatial dependence of the quasi-particle density of states is similar to the one with an $s$-wave order parameter, except possibly for the contribution from the extended states.

A further comment is in order on the existence of bound states in the continuum (usually referred to as BICs) that we have found in an isolated vortex for symmetry reasons, owing to the different values 
of the quantum $k_{z}$ between these bound states and the continuum states in which they are embedded.
As a consequence, a slight geometric perturbation of the vortex line along the $z$-direction could turn BIC states into decaying resonances \cite{FW-BIC-1985}. 
In this context, it is interesting to mention that BIC states have recently gained considerably experimental and technological interest as a generic wave phenomenon which can occur in supercavity lasing \cite{Kodigala-2017}.

On physical grounds, the Caroli-de~Gennes-Matricon bound states that are present in the interior of a vortex have a similar origin as the Andreev-Saint-James states that show up as subgap states in the context of 
the Josephson effect \cite{Andreev-1964,Saint-James-1964,Andreev-1965} (for a more recent review dealing also with cuprate superconductors, see Ref.~\cite{Deutscher-2005}).
Also for the Andreev-Saint-James states, a systematic theoretical study of their occurrence has been performed throughout the BCS-BEC crossover by solving the BdG equations at zero temperature \cite{SPS-2010}.
It would then be interesting to address experimentally the occurrence of these subgap states also along the way of the BCS-BEC crossover, in terms of suitable spectroscopic measurements \cite{Esteve-2013}.


\begin{center}
\begin{small}
{\bf ACKNOWLEDGMENTS}
\end{small}
\end{center}

This work was partially supported by the Italian MIUR under Contract PRIN-2015 No. 2015C5SEJJ001.

\appendix   
\section{INTERNAL STRUCTURE OF A VORTEX}
\label{sec:appendix}
\vspace{-0.2cm}

As mentioned in Section \ref{sec:formal_aspects}, the accurate numerical solution of the BdG equations performed in Ref.~\cite{SPS-2013} has evidenced the feature that the radial profile 
of the gap parameter has a ($\rho^{-2}$) power-law tail for \emph{all} couplings throughout the BCS-BEC crossover (as well as for \emph{all} temperatures in the superfluid phase).
In this Appendix we show that, in the BEC (strong-coupling) limit of the BCS-BEC crossover, whereby the fermionic BdG equations reduce to the bosonic Gross-Pitaevskii (GP) equation for the composite bosons that form in this limit \cite{Pieri-2003}, the ($\rho^{-2}$) long-range behavior of the condensate wave function $\Phi(\mathbf{r}) = \sqrt{\frac{m^{2} a_{F}}{8 \pi}} \, \Delta(\mathbf{r})$ can be determined by simple analytic considerations. 
Although this result has already been reported for a vortex filament in an almost ideal Bose gas described at low temperature by the GP equation \cite{LP-1961-2006}, the reason to briefly discuss it here is that its relevance for a vortex in a fermionic superfluid described by the BdG equations has passed essentially unnoticed in the literature \cite{Berthod-2017,Berthod-2016,Tempere-2017}.

The (time-independent) GP equation reads \cite{Pitaevskii_Stringari-2003}:
\begin{equation}
\left[ - \frac{1}{2 m_{B}} \nabla^{2} + V_{\mathrm{ext}}(\mathbf{r}) + \frac{4 \pi a_{B}}{m_{B}} |\Phi(\mathbf{r})|^{2} - \mu_{B} \right] \Phi(\mathbf{r}) = 0
\label{GP-equation}
\end{equation}
\noindent
where $m_{B}$, $a_{B}$, and $\mu_{B}$ are the bosonic mass, scattering length, and chemical potential, respectively.
[For composite bosons built up in terms of superfluid fermions described by the BdG equations, $m_{B}=2m$, $a_{B}=2a_{F}$, and $\mu_{B}=2\mu + \varepsilon_{0}$
where $\varepsilon_{0} = (m a_{F}^{2})^{-1}$ is the binding energy of the composite bosons \cite{Pieri-2003}.]

For an isolated vortex filament directed along the $z$ axis, one sets $V_{\mathrm{ext}}(\mathbf{r})=0$ and writes generically the vortex solution in cylindrical coordinates in the form 
$\Phi(\mathbf{r})= \sqrt{n_{0}} \, e^{i s \varphi} f(\rho)$, where $s$ is an integer referred to as the topological charge of the flow (we will set $s=1$ at the end of the calculation).
Far away from the vortex axis, one expects the local bosonic density $n(\mathbf{r})= |\Phi(\mathbf{r})|^{2}$ to reach its bulk value $n_{0}$, such that $\mu_{B}=U_{0} n_{0}$
where $U_{0}=\frac{4 \pi a_{B}}{m_{B}}$.
Introducing at this point the bosonic healing length $\xi_{B}= (2 m_{B} U_{0} n_{0})^{-1/2}$ and the rescaled radial variable $\eta = \rho/ \xi_{B}$, the GP equation (\ref{GP-equation}) acquires the form:
\begin{equation}
\frac{1}{\eta} \frac{d}{d \eta} \left( \eta \frac{d f(\eta)}{d \eta} \right) + \left( 1 - \frac{s^{2}}{\eta^{2}} \right) f(\eta) - f(\eta)^{3} = 0 \, .
\label{radial-GP-equation}
\end{equation}
\noindent
To determine the asymptotic behavior of $f(\eta)$ for $\eta \gg 1$, one sets
\begin{equation}
f(\eta) = 1 + C \eta^{\gamma} + \cdots
\label{asymptotic-behavior}
\end{equation}
\noindent
with $\gamma < 0$, and obtains the constants $C$ and $\gamma$ by entering the approximate form (\ref{asymptotic-behavior}) into Eq.~(\ref{radial-GP-equation}).
At the order here considered, one gets from Eq.~(\ref{radial-GP-equation}) the algebraic condition:
\begin{equation}
- s^{2} \eta^{-2} - 2 C \eta^{\gamma} + C (\gamma^{2} - s^{2}) \eta^{\gamma-2} - 3 C^{2} \eta^{2\gamma} - C^{3} \eta^{3 \gamma} = 0
\label{algebraic-condition}
\end{equation}
\noindent
which to the leading order yields $C=-s^{2}/2$ and $\gamma = -2$.
For $s=1$ it results that $f(\eta)$ tends to unity as $f(\eta) = 1 - \frac{1}{2\eta^{2}}$, as reported in Ref.~\cite{LP-1961-2006}.
Note that, physically, this long-range behavior is dominated by the angular kinetic energy of the vortex.




\begin{thebibliography}{99}

\bibitem{Berthod-2017} C. Berthod, I. Maggio-Aprile, J. Bru\'er, A. Erb, and C. Renner, \emph{Observation of Caroli-deGennes-Matricon vortex states in $YBa_{2}Cu_{3}O_{7-\delta}$}, Phys. Rev. Lett. {\bf 119}, 237001 (2017).

\bibitem{SX-2017} C.-L. Song and Q.-K. Xuey, \emph{Cuprate superconductors may be conventional after all}, Physics {\bf 10}, 129 (2017).

\bibitem{BCS-1957} J. Bardeen, L. N. Cooper, and J. R. Schrieffer, \emph{Theory of superconductivity}, Phys. Rev. {\bf 108}, 1175 (1957).

\bibitem{DeGennes-1966} P. G. de Gennes, \emph{Superconductivity of Metals and Alloys} (Benjamin, New York, 1966), Chap.~5.

\bibitem{Berthod-2016} C. Berthod, \emph{Vortex spectroscopy in the vortex glass: A real-space numerical approach}, Phys. Rev. B {\bf 94}, 184510 (2016).

\bibitem{CdGM-1964} C. Caroli, P. G. De Gennes, and J. Matricon, \emph{Bound fermion states on a vortex line in a type II superconductor},      
                                    Phys. Lett. {\bf 9}, 307 (1964).

\bibitem{Physics-Reports-2018} G. C. Strinati, P. Pieri, G. R\"{o}pke, P. Schuck, and M. Urban, \emph{The BCS-BEC crossover: 
                                                    From ultra-cold Fermi gases to nuclear systems}, Phys. Rep. {\bf 738}, 1 (2018).

\bibitem{Schluter-1991} F. Gygi and M. Schl\"{u}ter, \emph{Self-consistent electronic structure of a vortex line in a type-II superconductor},         
                                       Phys. Rev. B {\bf 43}, 7609 (1991).
                                       
\bibitem{Sensarma-2006} R. Sensarma, M. Randeria, and T-L. Ho, \emph{Vortices in superfluid Fermi gases through the BEC to BCS crossover}, Phys. Rev. Lett. {\bf 96}, 090403 (2006).    

\bibitem{Levin-2006} C.-C. Chien, Y. He, Q. Chen, and K. Levin, \emph{Ground-state description of a single vortex in an atomic Fermi gas: From BCS to Bose-Einstein condensation}, 
                                  Phys. Rev. A {\bf 73}, 041603(R) (2006).                           
 
\bibitem{SPS-2013} S. Simonucci, P. Pieri, and G. C. Strinati, \emph{Temperature dependence of a vortex in a superfluid Fermi gas}, 
                                 Phys. Rev. B {\bf 87}, 214507 (2013).       
                                 
\bibitem{Dagotto-1994} See, e.g., E. Dagotto, \emph{Correlated electrons in high-temperature superconductors}, Rev. Mod. Phys. {\bf 66}, 763 (1994).

\bibitem{RMP-2007} \O. Fischer, M. Kugler, I. Maggio-Aprile, C. Berthod, and C. Renner, \emph{Scanning tunneling spectroscopy of high-temperature superconductors}, Rev. Mod. Phys. {\bf 79}, 353 (2007).

\bibitem{Esteve-2008} H. le Sueur, P. Joyez, H. Pothier, C. Urbina, and D. Esteve, \emph{Phase controlled superconducting proximity effect probed by tunneling spectroscopy}, Phys. Rev. Lett. {\bf 100}, 197002 (2008).

\bibitem{Pitaevskii_Stringari-2003} L. Pitaevskii and S. Stringari, \emph{Bose-Einstein Condensation} (Clarendon Press, Oxford, 2003).

\bibitem{Pieri-2003} P. Pieri and G. C. Strinati, \emph{Derivation of the Gross-Pitaevskii equation for condensed bosons from the 
                                Bogoliubov-de~Gennes equations for superfluid fermions}, Phys. Rev. Lett. {\bf 91}, 030401 (2003).
                                
\bibitem{footnote} For continuum states above threshold, the quantum number $\nu_{\mathrm{r}}$ is replaced by the energy eigenvalue $\varepsilon$ and the $\sum_{\nu_{\mathrm{r}}}$ in
                             Eq.~(\ref{local-density_of_states}) is replaced by an integral over $\varepsilon$.
                                
\bibitem{Tempere-2017} N. Verhelst, S. Klimin, and J. Tempere, \emph{Verification of an analytic fit for the vortex core profile in superfluid Fermi gases}, Physica C: Superconductivity and its
                                       Applications {\bf 533}, 96 (2017). 
 
\bibitem{PPPS-2012} F. Palestini, A. Perali, P. Pieri, and G. C. Strinati, \emph{Dispersions, weights, and widths of the single-particle spectral function in the normal phase of a Fermi gas}, 
                                   Phys. Rev. B {\bf 85}, 024517 (2012).

\bibitem{Kato-2001} M. Kato and K. Maki, \emph{Bound states and extended states around vortices in $d$-wave superconductors}, J. Magn. Magn. Mater. {\bf 226-230}, 280 (2001).

\bibitem{FW-BIC-1985} H. Friedrich and D. Wintgen, \emph{Interfering resonances and bound states in the continuum}, Phys. Rev. A {\bf 32}, 3231 (1985).

\bibitem{Kodigala-2017} A. Kodigala, T. Lepetit, Q. Gu, B. Bahari, Y. Fainman, and B. Kant\'{e}, \emph{Lasing action from photonic bound states in continuum}, Nature {\bf 541}, 196 (2017).
                                
\bibitem{Andreev-1964} A. Andreev, \emph{The thermal conductivity of the intermediate state in superconductors}, Sov. Phys. JETP {\bf 19}, 1228 (1964) [Zh. Eksp. Teor. Fiz. {\bf 46}, 1823 (1964)].

\bibitem{Saint-James-1964} D. Saint-James, \emph{Elementary excitations in the vicinity of the surface separating a normal metal and a superconducting metal}, J. Phys. France {\bf 25}, 899 (1964).

\bibitem{Andreev-1965} A. Andreev, \emph{Electron spectrum of the intermediate state of superconductors}, Sov. Phys. JETP {\bf 22}, 455 (1966) [Zh. Eksp. Teor. Fiz. {\bf 49}, 655 (1965)].

\bibitem{Deutscher-2005} G. Deutscher, \emph{Andreev-Saint-James reflections: A probe of cuprate superconductors}, Rev. Mod. Phys. {\bf 77}, 109 (2005).

\bibitem{SPS-2010} A. Spuntarelli, P. Pieri, and G. C. Strinati, \emph{Solution of the Bogoliubov-de~Gennes equations at zero temperature 
                                  throughout the BCS-BEC crossover: Josephson and related effects}, Phys. Rep. {\bf 488}, 111 (2010). 

\bibitem{Esteve-2013} L. Bretheau, \c{C}. \"{O}. Girit, C. Urbina, D. Esteve, and H. Pothier, \emph{Supercurrent spectroscopy of Andreev states}, Phys. Rev. X {\bf 3}, 041034 (2013).

\bibitem{LP-1961-2006} L. P. Pitaevskii, \emph{Vortex lines in an imperfect Bose gas, }Sov. Phys. JETP {\bf 13}, 451 (1961) [Zh. Eksp. Teor. Fiz. {\bf 40}, 646 (1961)]; 
                                       see also, E. M. Lifshitz and L. P. Pitaevskii, \emph{Statistical Physics, Part 2, Theory of the Condensed State} (Butterworth-Heinemann, Oxford, 2006), Sec. 30.

\end{thebibliography}
\end{document}